\documentclass[aps,nobibnotes,nofootinbib,prl,reprint,groupedaddress]{revtex4-1}
\def\beq{\begin{eqnarray}}
\def\eeq{\end{eqnarray}}

\def\ba{\begin{eqnarray}}
\def\ea{\end{eqnarray}}
\def\beq{\begin{eqnarray}}
\def\eeq{\end{eqnarray}}

\def\mpl{M_{\rm Pl}}

\def\p{{\cal P}}

\def\L*{{\cal L}_*}
\def\L{\mathcal{L}}
\def\({\left(}
\def\){\right)}

\def\nn{\nonumber}
\def\p{\partial}

\def\p{\partial}

\def\<{\langle}
\def\>{\rangle}

\usepackage{hyperref}
\usepackage{amsmath}
\usepackage{amsfonts}
\usepackage{verbatim}
\usepackage{setspace}
\usepackage{url}
\usepackage{color}
\usepackage{subfig}
\usepackage{graphics}
\usepackage{graphicx}
\def\beq{\begin{eqnarray}}
\def\eeq{\end{eqnarray}}

\def\beq{\begin{eqnarray}}
\def\eeq{\end{eqnarray}}

\def\mpl{M_{\rm Pl}}

\def\p{{\cal P}}

\def\L*{{\cal L}_*}
\def\L{\mathcal{L}}
\def\({\left(}
\def\){\right)}

\def\nn{\nonumber}
\def\p{\partial}

\def\p{\partial}

\def\<{\langle}
\def\>{\rangle}

\def\beta{\beta}
\def\lsim{\mathrel{\rlap{\lower3pt\hbox{\hskip0pt$\sim$}}
     \raise1pt\hbox{$<$}}}         
\def\gsim{\mathrel{\rlap{\lower4pt\hbox{\hskip1pt$\sim$}}
     \raise1pt\hbox{$>$}}}         
\def\lsim{\mathrel{\rlap{\lower3pt\hbox{\hskip0pt$\sim$}}
     \raise1pt\hbox{$<$}}}         
\def\gsim{\mathrel{\rlap{\lower4pt\hbox{\hskip1pt$\sim$}}
     \raise1pt\hbox{$>$}}}         

\def\be{\begin{equation}}
\def\ee{\end{equation}}

\begin{document}


\title{}


\title{Large Non-Gaussianity in Slow-Roll Inflation}
\author{David Pirtskhalava}
\email[Electronic address: ]{david.pirtskhalava@sns.it} 
\affiliation{Scuola Normale Superiore, Piazza dei Cavalieri 7, 56126, Pisa, Italy}
\author{Luca Santoni}
\email[Electronic address: ]{luca.santoni@sns.it} 
\affiliation{Scuola Normale Superiore, Piazza dei Cavalieri 7, 56126, Pisa, Italy}
\affiliation{INFN -- Sezione di Pisa, 56200, Pisa, Italy}
\author{Enrico Trincherini}
\email[Electronic address: ]{enrico.trincherini@sns.it} 
\affiliation{Scuola Normale Superiore, Piazza dei Cavalieri 7, 56126, Pisa, Italy}
\affiliation{INFN -- Sezione di Pisa, 56200, Pisa, Italy}
\author{Filippo Vernizzi}
\email[Electronic address: ]{filippo.vernizzi@cea.fr} 
\affiliation{Institut de physique th\' eorique, Universit\'e  Paris Saclay}
\affiliation{CEA, CNRS, 91191 Gif-sur-Yvette, France}




\begin{abstract}

Canonical models of single-field, slow-roll inflation do not lead to appreciable non-Gaussianity, unless derivative interactions of the inflaton become uncontrollably large. 
We propose a novel slow-roll scenario  where scalar perturbations propagate at a subluminal speed, leading to sizeable equilateral non-Gaussianity, $f^{\rm equil}_{\rm NL}\propto 1/c_s^4$, largely insensitive to the ultraviolet physics. The model is based on a low-energy effective theory characterized by weakly broken invariance under internal galileon transformations, $\phi\to\phi+b_\mu x^\mu$, which protects the properties of perturbations from large quantum corrections. This provides the unique alternative to models such as DBI inflation in generating strongly subluminal/non-Gaussian scalar perturbations.  
 
\end{abstract}


\maketitle

\paragraph{ \it \bf Introduction:}
The simplest, textbook version of inflation consists of a single canonical scalar field---the inflaton---slowly rolling down a sufficiently flat potential. 
It is a common feature of these models that the magnitude of non-Gaussianity is suppressed  \cite{Maldacena:2002vr} well below the observable level for any foreseeable future; see, e.g.,~\cite{Alvarez:2014vva}.

Generically, in the low-energy inflationary effective field theory (EFT),   there are additional higher-derivative corrections to the canonical action, and one may wonder whether these can impact non-Gaussianity in any significant way. For example, consider the following Lagrangian 
\beq
\label{dphi4}
\mathcal{L}=\sqrt{-g}\bigg[-\frac12 (\partial \phi)^2 - V(\phi)+\frac{(\p\phi)^4}{\Lambda^4}+\dots ~\bigg],
\eeq
where $V(\phi)$ denotes the inflaton potential and the last term provides a higher-order correction in the derivative expansion, governed by the EFT cutoff $\Lambda$. The virtue of such terms is that they do not renormalize the potential, while they do contribute to the non-Gaussianity \cite{Creminelli:2003iq},
\beq
\label{fnl1}
f_{\rm NL}\sim \frac{\dot{\phi_0}^2 }{\Lambda^4}~,
\eeq
where $\phi_0$ denotes the background expectation value of $\phi$ and $f_{\rm NL}$ is the nonlinearity parameter of scalar perturbations \cite{Ade:2015ava}. It is clear from eq.~\eqref{fnl1} that having $f_{\rm NL}\gsim 1$ in this model implies going beyond the (at least apparent) regime of validity of the low-energy effective theory, which requires $(\partial \phi)^2\ll \Lambda^4$. Therefore, unless the infinite number of derivative operators in the ellipses of eq. \eqref{dphi4} can be resummed, one cannot trust values of non-Gaussianity greater than one. An example of such a resummation is provided by Dirac-Born-Infeld (DBI) inflation \cite{Silverstein:2003hf,Alishahiha:2004eh}, where 
a higher-dimensional spacetime symmetry (nonlinearly realized on $\phi$) protects the coefficients of the leading derivative operators from large quantum corrections. For a small speed of sound of scalar perturbations, $c_s^2\ll 1$, DBI inflation predicts equilateral non-Gaussianity \cite{Creminelli:2005hu}, with the amplitude  $f^{\rm equil}_{\rm NL}\sim 1/c_s^2$ \cite{Alishahiha:2004eh,Chen:2006nt}.

In this Letter, we propose a novel inflationary scenario, where the energy density of the early universe is dominated by the potential of a slowly-rolling scalar field, similarly to ordinary slow-roll theories. Yet, a definite set of higher-derivative interactions of the inflaton become relevant, leading to observably large non-Gaussianities. Nevertheless, the theory is predictive, since all the rest of the operators in the derivative expansion remain {\em naturally} small in the full quantum theory. 
These properties follow from the \textit{weakly broken} \cite{Pirtskhalava:2015nla} invariance under  internal \textit{galileon transformations} \cite{Nicolis:2008in},
\beq
\label{gi}
\phi\to \phi+b_\mu x^\mu~,
\eeq
that defines the Lagrangian our theory is based on.
While the symmetry \eqref{gi} has appeared in a variety of physical contexts, ranging from modified gravity \cite{Dvali:2000hr,Nicolis:2008in,deRham:2010ik,deRham:2010kj} to scattering amplitudes
\cite{Cheung:2014dqa}, here we use it as a guideline for constructing largely UV-insensitive models of the early universe. To our knowledge, \eqref{gi} is the only alternative to DBI-like symmetries for protecting the coefficients of strongly-coupled higher-derivative operators against large quantum corrections.

Just like in DBI inflation, enhanced scalar non-Gaussianity is generically associated with a reduced speed of sound of perturbations in our model;
however, the enhancement is much stronger compared to the DBI case, the amplitude of equilateral non-Gaussianity growing as $f^{\rm equil}_{\rm NL}\propto 1/c_s^4$ for small $c_s^2$. 

\paragraph{ \it \bf The model:} The theory we wish to study below is defined as a combination of an inflationary potential and the four Lagrangian terms \cite{Pirtskhalava:2015nla}, which, barring the factors of $\sqrt{-g}$, can be written as follows 
\begin{align}
L_2&=\Lambda_2^4 G_2 (X) \;, \label{L2} \\ 
L_3&={\Lambda_2^4} G_3 (X)   [\Phi]  \;, \label{L3}\\
L_4&= \mpl^2  G_{4} (X)  R+2 {\Lambda_2^4}  G_{4}' (X)   \left( [\Phi]^2- [\Phi^2]  \right)\;, \label{L4}\\
L_5&=  \mpl^2 G_{5} (X)  G_{\mu\nu}\Phi^{\mu\nu} \nonumber \\
&-\frac{1}{3} {\Lambda_2^4}  G_{5}' (X)   \left( [\Phi]^3-3 [\Phi][\Phi^2] +2 [\Phi^3] \right) \label{L5} \;.
\end{align}
Here, $\Phi$ is a matrix, consisting of second derivatives of the inflaton, 
$\Phi^\mu_\nu \equiv {\nabla^\mu \nabla_\nu \phi}/{\Lambda_3^3 }$, 
and the brackets $[ \ldots]$ denote the trace operator. Moreover, $G_a$ are arbitrary dimensionless functions of the dimensionless variable\footnote{We use the `mostly plus' signature for the metric.} 
$X\equiv -{g^{\mu\nu}\nabla_\mu\phi\nabla_\nu\phi}/{\Lambda_2^4} $ .
For simplicity, we will assume that these functions can be Taylor-expanded around zero, $G_a (X) = \sum_{n=0}^\infty c^{(n)}_{a} X^n$.
Furthermore, the two scales in the theory are related to each other as $\Lambda_2^4 = \mpl \Lambda_3^3$, so that the smaller of these, $\Lambda_3$, can be regarded as the genuine cutoff of the underlying low-energy EFT (we assume $\mpl$ is the parametrically highest scale in the problem).

In the limit $\mpl\to\infty$, the  Lagrangian terms   in  \eqref{L2}--\eqref{L5} reduce to the \textit{galileons} of Ref. \cite{Nicolis:2008in}, which are exactly invariant (up to a total derivative) under \eqref{gi}. In this limit, there is a non-renormalization theorem, according to which the galileon operators are not corrected, at least perturbatively, by quantum loops \cite{Luty:2003vm}. 
For a finite Planck mass, the operators \eqref{L2}--\eqref{L5} break the galileon symmetry, but only \textit{weakly} \cite{Pirtskhalava:2015nla}. The defining property of the theories with weakly broken invariance under \eqref{gi} is a generalization of the non-renormalization theorem of Ref. \cite{Luty:2003vm}, which renders the quantum corrections to the coefficients $c^{(n)}_{a}$  suppressed by positive (integer) powers of the tiny ratio $\Lambda_3/\mpl$ \cite{Pirtskhalava:2015nla}. In addition, these theories belong to the so-called Horndeski class of scalar-tensor models \cite{Horndeski:1974wa}, characterized by second-order equations of motion both for the scalar and  the metric \cite{Deffayet:2009mn}.


It has been shown in Ref.~\cite{Pirtskhalava:2015nla} that the properties of the theories with weakly broken galileon symmetry imply the possibility of a moderately coupled, yet predictive, regime characterized by
\beq
\label{stability}
X=\frac{\dot\phi_0^2}{\Lambda_2^4}\lsim 1 ~, \qquad Z\equiv \frac{H \dot\phi_0}{\Lambda_3^3}\lsim 1~,
\eeq
for a homogeneous $\phi$-profile on a FRW background with the Hubble rate $H$. From now on, $X$ will be understood as evaluated on the background solution. 
Despite the  moderate coupling, quantum corrections are under control when the scalar background profile satisfies \eqref{stability}---even in the case that these inequalities are saturated---and the predictions of the classical theory can be trusted. 

\paragraph{ \it \bf The slow-roll backgrounds:} 

As noted above, we will be interested in the potentially-dominated models of inflation, characterized by weakly broken invariance under the galileon transformations. These are governed by the following Lagrangian
\be
\label{full}
\mathcal{L}=\sqrt{-g} \bigg[\frac{\mpl^2}{2} R -\frac{1}{2}(\p\phi)^2 -V(\phi)+\sum_{i=2}^5 L_i +\dots\bigg] \;,
\ee
where, since we have extracted the  canonical scalar and graviton kinetic terms, $G_2$ is assumed to start at least quadratic in $X$, while $G_3$ can have a linear piece. From now on we will set $G_4 = G_5 =0$ for the sake of simplicity;  generalization to the case of nonzero $G_4$ and $G_5$ is straightforward and will be commented on where appropriate. The ellipses in \eqref{full} denote an infinite number of other operators, present in the low-energy effective theory. 
We will assume that  the potential  $V(\phi)$ satisfies the ordinary slow-roll conditions, $\epsilon_V \ll 1$ and $|\eta_V | \ll 1$, where the (potential-based) slow-roll parameters $\epsilon_V$ and $\eta_V$ are defined as
\be
\epsilon_V \equiv  \frac{\mpl^2}{2} \left( \frac{V'}{V} \right)^2 \;,  \qquad \eta_V \equiv \mpl^2 \frac{V''}{V}\;. \label{srp}
\ee
The previous analysis of quantum loops, leading to the non-renormalization theorem summarized above, has concentrated on the case with a vanishing potential \cite{Pirtskhalava:2015nla}. 
It is straightforward to show that the same results remain intact also in the presence of a nonzero, but sufficiently flat $V(\phi)$, satisfying \eqref{srp}.

For the flat FRW ansatz, $ds^2 = -dt^2 + a^2(t) d \vec x^2$,
the two Friedmann equations that follow from \eqref{full} read
\begin{align}
\label{fried1}
3\mpl^2H^2&=V - \Lambda_2^4 X\bigg[ \frac{1}{2}+ \frac{G_2}{X}- F(X,Z) \bigg], \\
\label{fried2}
2\mpl^2\dot H& =-\Lambda_2^4 X F(X,Z) +2\mpl X G'_{3} \ddot \phi_0~,
\end{align}
where $H \equiv \dot a/a$ and the function $F(X,Z)\equiv 1+2 G'_{2}-6 Z G'_{3}$ has been introduced for later convenience.
Moreover, in the slow-roll regime the homogeneous equation of motion of $\phi$ reduces to
\beq
\label{phiequation}
3 H \dot\phi_0 F(X,Z) \simeq -V'(\phi_0)~.
\eeq
We are interested in a regime where higher-derivative operators in \eqref{full} become important, while the quantum corrections are still under control. To this end, we assume  $Z\sim 1$, which also fixes the magnitude of the parameter $X$. Indeed, from the definition of $X$ and $Z$, eq.~\eqref{stability}, it follows that $\sqrt{X}=\Lambda_2^2 Z/(\mpl H)$; making use of eq. \eqref{fried2}, one immediately obtains $X\sim \sqrt{\epsilon}$, where $\epsilon \equiv - \dot H /H^2$.

Note the order-unity slowly varying function of time $F(X,Z)$ in eqs.~\eqref{fried1}--\eqref{phiequation}, which is strictly one in canonical slow-roll inflation. Apart from this minor modification, all the equations that describe the background solution are similar to those of ordinary slow-roll models (up to corrections of higher order in $\epsilon_V$ and $\eta_V$). In particular, unlike e.g. the DBI case, the usual flatness conditions $\epsilon_V \ll 1$ and $|\eta_V | \ll 1$ need to hold for sustaining the quasi-de Sitter phase in our model. It is precisely for this reason that we refer to it as ``slow-roll''.\footnote{Inflationary models based on particular subsets of the Lagrangian terms in \eqref{L2}--\eqref{L5} have been studied in Refs. \cite{Kobayashi:2010cm, Burrage:2010cu}. However, these references have focused on kinetically-driven inflation, corresponding to $\Lambda_2^4 \sim \mpl^2 H^2$ and $X \sim Z\ \sim 1$ in our notation.} 
 At the level of perturbations, our scenario is of course very different from the canonical slow-roll inflation; for example, unlike the latter, the scalar perturbations become strongly coupled at an energy scale parametrically smaller than $\mpl$, something we discuss in greater detail below.
 
 It follows from the Friedmann equations that the contributions from the derivative operators in \eqref{full} to the inflationary energy density and pressure are proportional to $X\sim \sqrt{\epsilon}$. One may wonder therefore, whether loop corrections can outweigh these contributions for small values of $\epsilon$. For $Z\sim 1$, the leading quantum corrections to the background stress tensor scale as $\sim \Lambda_3^4$ \cite{Pirtskhalava:2015nla}. This should be much smaller than $\Lambda_2^4 \sqrt{\epsilon}$, which implies a lower bound on the slow-roll parameter, $\epsilon \gg\(H/\mpl\)^2$. This is  the same bound on $\epsilon$ as the one that arises from requiring quantum fluctuations of the inflaton to be small \cite{Linde:1986fd,Goncharov:1987ir}. 

\paragraph{ \it \bf Non-Gaussianity:} 
Inflationary theories can be conveniently studied  in a model-independent way using the EFT framework \cite{Creminelli:2006xe,Cheung:2007st}. To this end, we decompose the metric in the ADM variables,
\be
ds^2=-N^2 dt^2+\gamma_{ij}\(N^i dt+dx^i\)\(N^j dt+dx^j\) \;,
\ee
and work in the unitary gauge, where the constant-time hypersurfaces are chosen to coincide with those of uniform $\phi$. The perturbed quantities are defined as $\delta N \equiv N-1$, 
and $\delta K \equiv K - 3H$, where $K$ denotes the trace of the extrinsic curvature of equal-time hypersurfaces. The action in eq. \eqref{full} (with $G_4 = G_5=0$) can be expanded to the cubic order in perturbations in the following way \cite{Gubitosi:2012hu}
\be
\begin{split}
\label{s_pi}
S &=\int d^4x\sqrt{ -g}\bigg[ \frac{\mpl^2}{2} R -\mpl^2 \dot H \frac{1}{N^2} -\mpl^2 (3 H^2+\dot H)  \\
&+\frac{M_2^4}{2}  \delta N^2   +M^4_3\delta N^3 -\hat M_1^3 \delta K \delta N +\hat M_2^3  \delta K \delta N^2  \bigg] \;,
\end{split}
\ee
where $M_2$, $M_3$, $\hat M_1$ and $\hat M_2$  are functions of time of canonical dimension one.
In terms of the functions $G_2$ and $G_3$ introduced in eqs.~\eqref{L2} and \eqref{L3}, these are 
\be
\begin{split}
\label{M4}
M_2^4 &= -2\Lambda_2^4 X\bigg[ 3Z G_{3}' +6 Z X G_{3}''-2 X G_{2}''-Y G_{3}'\bigg] \;, \\
M_3^4 &=-2\Lambda_2^4 X \bigg[3 X G_{2}''+\frac{2}{3} X^2 G_{2}''' \\
&\quad -Z \(4 G_{3}'+11 XG_{3}''+2 X^2G_{3}'''\)+Y G_{3}'\bigg] \;, \\
\hat M_1^3 &= -2\frac{\Lambda_2^4 X}{H} Z G_{3}' \, ,\quad \hat M_2^3 =-2\frac{\Lambda_2^4 X}{H}Z\(2 G_{3}'+X G_{3}''\) \;,
\end{split}
\ee
where $Y \equiv \ddot \phi_0/\Lambda_3^3$.
One can see from \eqref{M4}, that approximate invariance under galileon transformations imposes the following (radiatively stable) hierarchy among the various EFT coefficients:
\beq
\label{eftmagnitudes}
M_2^4\sim M_3^4\sim \mpl^2 \dot H, \qquad \hat M_1^3\sim \hat M_2^3\sim \frac{\mpl^2 \dot H}{H}~. \;~~~
\eeq 
This is in stark contrast to what happens e.g.~in solely shift-symmetric theories, where the coefficients that stem from higher-derivative operators such as $\hat M_1^3$ and $\hat M_2^3$, are much stronger suppressed. The latter hierarchy motivates to define the dimensionless, order-one coefficients 
\be
\label{alphas}
\begin{split}
\alpha_{1,3} &\equiv - \frac{M_{2,3}^4}{2 \mpl^2 \dot H}\;, \quad  \alpha_{2,4} \equiv - \frac{\hat M_{1,2}^3 H }{2  \mpl^2 \dot H} \;,
\end{split}
\ee
convenient for describing the parameter space of the theories at hand. 

At sufficiently high energies (encompassing the Hubble scale), the dynamics of scalar perturbations is fully dominated by the dynamics of the adiabatic mode $\pi$, defined through $\phi(\vec{x},t)=\phi_0\(t+\pi(\vec{x},t)\)$ \cite{Creminelli:2006xe,Cheung:2007st}. In the \textit{decoupling limit} corresponding to this regime, the scalar part of the action \eqref{s_pi} reads
\be
\begin{split}
\label{dlaction}
S_\pi &= \int d^4 x~a^3~(-\mpl^2 \dot H) \bigg[ (1+\alpha_1)\(\dot\pi^2-c_s^2 \frac{(\p\pi)^2}{a^2}\)   \\
&+ \left(  {\alpha_2}-\alpha_1 \right) \dot\pi\frac{(\p\pi)^2}{a^2} - 2 (  \alpha_1 + \alpha_3) \dot\pi^3 \\
&
+ 2 \frac{ \alpha_2 - \alpha_4}{H} ~\dot\pi^2 \frac{\p^2\pi}{a^2} + \frac{\alpha_2}{H} ~\frac{(\p\pi)^2\p^2\pi}{a^4}  \bigg]~,
\end{split}
\ee
where the speed of sound is
\beq
\label{cssq}
c_s^2 \equiv  \frac{1 + \alpha_2}{1 + \alpha_1}  \;.
\eeq
It follows from eq. \eqref{cssq} that if, for whatever reason, the parameter $\alpha_2$ happens to be close to $-1$, one can have strongly subluminal scalar perturbations. Most importantly, the approximate galileon invariance guarantees that such an `accidental' arrangement of the parameters is respected by loop corrections. 
This is qualitatively different from how the small $c_s^2$ arises in models such as DBI inflation, as we discuss in detail below\footnote{In the limit $X\to 0$, using eq.~\eqref{M4} with $G_4 = G_5=0$ in eq. \eqref{alphas}, one finds $\alpha_1= 3\alpha_2$. This gives a negative kinetic term to $\pi$ for $\alpha_2 \simeq -1$. The parameter $X$ need not be very small, however; it is of order $X \sim \sqrt{\epsilon}\simeq \text{a few}\times 0.1$  in e.g. slow-roll models with monomial potentials. Moreover, the relation  $\alpha_1 = 3\alpha_2$ no longer holds for $G_4 \neq 0$ or $G_5 \neq 0$.}.

It is worth stressing at this point that the operators in the last line of \eqref{dlaction} can be  rewritten in terms of those in the second line via a perturbative field redefinition \cite{Creminelli:2010qf}. 
This simply amounts to using the linear equation of motion in \eqref{dlaction}. After straightforward manipulations, one finds\footnote{Explicitly, the coefficients $\gamma_1$ and $\gamma_2$ read
\begin{align}
\gamma_1 &\equiv  (c_s^2 - 1) \left(1+\frac{2}{c_s^2}\right) + (1+c_s^2)\alpha_1 \;, \nn\\
\gamma_2  &\equiv 2 \left(1- \frac1{c_s^{2}} \right) \left(2+ \frac1{c_s^{2}}\right) + \frac{2}{c_s^2} (  \alpha_1 - 2 \alpha_4) + 2 \alpha_1 - 2 \alpha_3 
\;.\nn
\end{align}} 
\be
\label{dbiops}
S^{(3)}_{\pi} = \int d^4 x ~a^3~ (- \mpl^2 \dot H) \bigg[\gamma_1 \dot\pi\frac{(\p\pi)^2}{a^2} + \gamma_2  \dot\pi^3  \bigg] \;.
\ee
The two operators in \eqref{dbiops} are precisely those appearing in the decoupling limit of DBI theories and the bispectrum they produce is close to the equilateral shape \cite{Creminelli:2005hu}. The genuine difference arises once the \textit{magnitude} of non-Gaussianities is concerned: instead of the $f^{\rm equil}_{\rm NL}\sim 1/c_s^2$ behaviour characteristic of DBI inflation \cite{Chen:2006nt},  in theories with weakly broken galileon symmetry non-Gaussianity scales as $f^{\rm equil}_{\rm NL}\sim 1/c_s^4$ in the small-$c_s^2$ limit. The latter scaling is due to the last operator in \eqref{dlaction}, whose precise contribution to the three-point function of the curvature perturbation $\zeta$ reads \cite{Mizuno:2010ag,Burrage:2010cu}\footnote{We follow the standard definition of the comoving curvature perturbation, $g_{ij}=a^2 e^{2\zeta}\delta_{ij}$. The three-point function is defined as $\langle \zeta(\vec k_1)\zeta(\vec k_2)\zeta(\vec k_3)\rangle=(2\pi)^3\delta^{(3)}\(\sum_i \vec{k}_i\)B_\zeta(k_1,k_2,k_3)$, and $k_t\equiv k_1+k_2+k_3$~. }
\be
\label{bzeta}
\begin{split}
B_\zeta&(k_1,k_2,k_3) =-\frac{1}{16}\frac{H^8}{A^2}\frac{\alpha_2}{1+\alpha_1}\frac{1}{c_s^{10}} \frac{k_1^2(k_1^2-k_2^2-k_3^2)}{k_t (k_1k_2k_3)^3} \\
&\times \(1+\frac{ \sum_{i>j}k_ik_j}{k_t^2}+3\frac{k_1k_2k_3}{k_t^3}\)+2~\text{perms}~.
\end{split}
\ee
Here,  $A$ denotes the normalization of the $\pi$-kinetic term in the decoupling limit, $A\equiv (-\mpl^2 \dot H)(1+\alpha_1)$. 
The amplitude of non-Gaussianity can be directly read off from eq.~\eqref{bzeta}
\beq
\label{fnl}
f^{\rm equil}_{\rm NL}=\frac{5}{18}\frac{k_*^6 B_\zeta(k_*,k_*,k_*)}{\Delta^2_{\zeta*}}=\frac{65}{162}\frac{\alpha_2}{1+\alpha_1}\frac{1}{c_s^4}~,
\eeq
where $\Delta_\zeta \equiv  k^3 P_{\zeta} (k) = H^4/(4Ac_s^3)$ is the dimensionless power spectrum, evaluated at a fiducial momentum scale $k_*$. A significantly reduced speed of sound, $\alpha_2 \simeq -1$ (see eq.~\eqref{cssq}), implies a \textit{negative} $f_{\rm NL}$ . However, due to the strong dependence of $f_{\rm NL}$ on $c^2_s$, even slightly subluminal perturbations can produce a sizeable amount of non-Gaussianity; for example, a $ 10~ \%$ tuning of the $\alpha_2$ parameter can give rise to $f^{\rm equil}_{\rm NL}\simeq -70$ for $\alpha_1=1$ and $\alpha_2=-0.9$. We stress that such a tuning is not `unnatural' as a result of the non-renormalization theorem outlined above.

If the theory \eqref{dlaction} is to be predictive, it is crucial that $\pi$ is weakly coupled at energies of the order of the inflationary Hubble rate, $\Lambda_\star\gg H$, where $\Lambda_\star$ is the energy scale at which perturbative unitarity is violated in the $2\to2$ scattering of $\pi$. In the $c_s^2\ll 1$ limit, $\Lambda_\star$ is set by the last interaction term in the action \eqref{dlaction}, and is estimated as $\Lambda_\star\sim \Lambda_3  c_s^{11/6}$. For $\alpha_1 \simeq 1$ and $\alpha_2 \simeq -1$,  using the experimental value $\Delta_\zeta \simeq 5 \pi^2 \times 10^{-9}$,  one finds
\be
\Lambda_\star^3 \sim ~\frac{\mathcal{O}(50)}{\big| {f_{\rm NL}^{\rm equil}} \big| }~(8 H)^3 \;.
\ee
Even for the largest $f_{\rm NL}^{\rm equil}$ compatible with the current observational bounds \cite{Ade:2015ava}, the strong coupling scale is fairly above $H$, but well below the symmetry breaking scale  \cite{Baumann:2011su}, $\Lambda_b^3 \sim {\cal O}(5) \big| {f_{\rm NL}^{\rm equil}} \big| \Lambda^3_\star $.


We close this section with a remark concerning the regime of validity of the decoupling limit. For the small values of the speed of sound we are interested in, one should be  careful with mixing terms that involve spatial derivatives. For example, consider the $\delta N\delta K$ operator in eq. \eqref{s_pi}.  The most important mixing of scalar modes with gravity that arises from this operator is suppressed by a factor of $\epsilon/c_s^2$ compared to the $\pi$-kinetic term at horizon crossing. Therefore, the validity of the decoupling limit analysis requires that $ c_s^2 \gsim \epsilon$ hold, which puts an upper bound ($f_{\rm NL}\lsim 1/\epsilon^2$) on the amplitude of non-Gaussianity attainable within the decoupling limit. 

\paragraph{ \it \bf Discussion:} 

In `ordinary' inflationary  theories, the statistical properties of perturbations are mostly determined by operators with the least number of derivatives, i.e.~$(\p\phi)^{2n}$ or, in the EFT language, by operators of the form $\(\delta N\)^n$. The prototype example of this is provided by DBI inflation, where the background can be consistently strongly coupled, with an infinite number of operators of the above type becoming relevant for its dynamics. In the EFT of eq. \eqref{s_pi}, this corresponds to the large coefficient $M_2^4$ or, equivalently, to $\alpha_1\gg 1$ in the notation of eq.~\eqref{alphas}, and implies a parametrically suppressed speed of sound, $c_s^2 \simeq  1/\alpha_1$, see eq.~\eqref{cssq} (higher-derivative operators are effectively negligible in DBI inflation, so one can set $\alpha_2 \simeq 0$). Despite the strong coupling, the symmetries of DBI theories protect the structure of the Lagrangian from large quantum corrections. 
This mechanism of obtaining strongly sub-luminal scalar perturbations has been extensively studied in the literature (see, e.g., \cite{Alishahiha:2004eh,Chen:2006nt,Senatore:2009gt,Baumann:2011su,Baumann:2014cja}) and provides an attractive way of generating large equilateral non-Gaussianity in single-field inflation.

In this work we have proposed an alternative scenario that allows for strongly subluminal scalar perturbations within a well-defined low-energy EFT. 
Our model relies on moderate coupling, i.e.~$\alpha_i\sim 1$, as a result of which more than one operator in  eq.~\eqref{s_pi} become
large enough to affect the scalar perturbations significantly.
In the simplest realization considered above, the scalar two- and three-point functions are determined by the operators $\delta N^2$ and $\delta N\delta K$. Instead of the $\alpha_1\gg1,~ \alpha_2\ll 1$ case characteristic of DBI inflation, our mechanism relies on an adjustment of the order-unity $\alpha_1$ and $\alpha_2$, which results in a somewhat reduced speed of sound in \eqref{cssq}.
The theories we have studied in this work not only allow for such an adjustment, but also provide a way to protect it against loop corrections. The central reason behind the robustness of the results obtained in the classical theory is the \textit{weakly broken} galileon symmetry \cite{Pirtskhalava:2015nla}, which inherits the remarkable quantum properties of the galileon operators \cite{Nicolis:2008in}, exactly invariant under \eqref{gi}. 

\paragraph{ \it \bf Acknowledgements:} 
We thank Daniel Baumann, Paolo Creminelli and Greg Gabadadze for valuable discussions and comments on the manuscript. The work of D.P. is supported in part by MIUR-FIRB grant RBFR12H1MW and by funds provided by {\it Scuola Normale Superiore} through the program {\it "Progetti di Ricerca per Giovani Ricercatori"}. The work of E.T. is supported in part by MIUR-FIRB grant RBFR12H1MW. F.V. acknowledges partial support from the grant ANR-12-BS05-0002 of the French Agence Nationale de la Recherche.

\bibliography{eftinf}

\end{document}